\documentstyle[preprint,aps]{revtex}
\tightenlines
\begin{document}
\draft
\preprint{\vbox{Submitted to Physical Review {\bf C}\hfill
        USC(NT)-98-05}}
\tolerance = 10000
\hfuzz=5pt
\title{Pion-nucleon scattering and the nucleon sigma term in an
extended linear sigma model }
\author{V. Dmitra\v sinovi\' c, and F. Myhrer}
\address{Department of Physics  and Astronomy,\\
University of South Carolina, Columbia, SC 29208, USA} 
\date{\today}
\maketitle
\begin{abstract}
A modified linear sigma model that allows 
for $g_{A} = 1.26$ by addition of vector and pseudovector $\pi N$ coupling 
terms was discussed by Bjorken and Nauenberg and by Lee. %\cite{bn68}.
In this extended linear sigma model 
the elastic $\pi N$ scattering amplitudes satisfy 
the relevant chiral low-energy theorems, such as the 
Weinberg-Tomozawa relation for the isovector $\pi N$ 
scattering length and in some cases Adler's ``consistency condition''. 
The agreement of the isospin symmetric $\pi N$ scattering length 
with experiment is substantially 
improved in this extended sigma model as compared 
with the original linear one. 
We show that the nucleon sigma term ($\Sigma_N$)
in the linear- and the extended sigma models 
with three different kinds of
chiral symmetry breaking terms are identical. 
Within the tree approximations  the
formal operator expression for the $\Sigma_N$ term and 
the value extracted from
the $\pi$N scattering matrix coincide.  
Large values of $\Sigma_N$ are easily obtained without any  
$s \bar{s}$ content of the nucleon. 
Using chiral rotations the 
Lagrangian of this extended sigma model 
reproduces the lowest-order $\pi$N chiral perturbation theory
Lagrangian.
\end{abstract}
%-------------------------------------------------------------------------------
\widetext

\section{Introduction}
Gell-Mann and Levy's [GML]
linear sigma model is a principal example of 
spontaneously broken chiral symmetry in 
strong interactions\cite{gml60}. 
It is known that the linear sigma model does not always give 
the correct phenomenology, e.g. the value of the isoscalar pion-nucleon
scattering length is too large. 
We shall show that in the extended linear sigma model to be presented 
in this paper, the phenomenology is considerably
improved compared to the original GML model.
Another alleged drawback of the linear sigma model is that, apart from 
chiral symmetry, the model has not been connected directly to QCD. 
Recently, however, it has been shown that 
%
%The linear sigma model has often been maligned as: (1) not 
%having a basis in QCD; (2) making incorrect phenomenological predictions.
%The first objection has been met in Ref. \cite{u196}, where it was 
%shown that 
%
the model can be thought of as a low-energy effective theory of 
Coulomb gauge QCD, albeit in the unrealistic limit of maximal 
$U_{A}(1)$ symmetry breaking\cite{u196}. 
%
%In other words, so long as one does not
%address $U_{A}(1)$ symmetry, the linear $\sigma$ model is as good as 
%Coulomb gauge QCD, or any other chiral model. 
%And if one wishes to address the $U_{A}(1)$ question, 
%then one must extend the model to include all relevant meson fields, i.e.,
%one must work with the `t Hooft model for $N_{f}=2$, or with L\' evy's
%model for $N_{f}=3$ \cite{u196}.
%On the second charge, 

%Yet a
Another ``weakness" of the linear sigma model is that the value of 
the axial coupling strength $g_A$ equals one. 
It is known that 
the one-loop ``radiative'' corrections 
in the linear sigma model lead to the 
renormalization of the nucleon part of the axial current \cite{peris92}, 
but it is not widely known how to 
incorporate that kind of correction, i.e. a value of $g_A \ne 1$, 
into an effective (tree-level) Lagrangian. 
In some publications a proposed  ``solution'' is 
to multiply  the total axial current 
${\bf J}_{\mu 5}^{a} = {\bf A}_{\mu}^{a} + {\bf a}_{\mu}^{a}$ by $g_A$  
where the nucleon part 
of the axial current is 
${\bf A}_{\mu}^{a} =
\bar{\psi}\gamma_{\mu} \gamma_{5} {\bbox{\tau}^{a}\over 2}\psi$, and 
the meson part of the axial current is 
${\bf a}_{\mu}^{a} =  \sigma \partial_{\mu} \bbox{\pi}^{a} - 
\bbox{\pi}^{a} \partial_{\mu} \sigma$. 
Another ``solution'' posits the same, but this time just for  ${\bf A}_{\mu}^{a}$.
Both of these ``solutions" are inconsistent with the chiral symmetry 
of the model. The first one violates the chiral charge algebra by leading to
\begin{eqnarray}
\left[ Q_{5}^{a} , Q_{5}^{b}\right] 
&=&  g_A^2 {\rm i} \varepsilon^{abc} Q^{c}
\neq {\rm i} \varepsilon^{abc} Q^{c}
~. \label{e:axcr} \
\end{eqnarray}
The second ``solution" leads to Eq.(\ref{e:axcr})  for 
the nucleon part of the axial charge, and in addition to 
a non-conserved axial Noether current even in the
chiral limit since  the 
equations of motion have {\it not} been modified. 
 
In an earlier publication \cite{ax96} one of us 
re-initiated the study of a venerable, but little-known 
%modification or 
extension of the linear sigma model,
%due to Bjorken and Nauenberg 
see e.g. Ref.\cite{bn68}. 
This extension  
allows the nucleon axial coupling constant $g_A$
to be different from unity without violating chiral symmetry. 
The extra term introduced in  the linear sigma model 
is a non-renormalizable, derivative-coupling term, 
analogous to the Pauli anomalous (electron) magnetic moment term 
that describes the finite
one-loop radiative correction in QED, and 
that is often introduced into other effective Lagrangians.
This extended linear sigma model allows one to study the
$g_A$ dependence of the $\pi$N scattering lengths, $a_{\pi N}$,  
and of the nucleon sigma term $\Sigma_N$.
It is well known that  $a_{\pi N}^{(-)}$ depends crucially 
on the value of $g_A$, 
%[Adler-Weisberger, or Weinberg-Tomozawa result], 
whereas the  $\Sigma_N$ dependence on $g_A$ is unknown \cite{camp79}. 
We shall display this dependence and show that a large value of 
$\Sigma_N$ can easily be obtained 
without recourse to any $s\bar{s}$ component of the nucleon. 
We can also reproduce the new, tiny experimental 
value of the isoscalar $\pi$N scattering length $a_{0}^{(+)}$. 
Our methods and results are potentially important for studies
of nuclear matter, because the quark condensate in nuclear matter is 
determined by the n-nucleon sigma terms \cite{vd99,delor96}, and
the issue of (P-wave) pion condensation depends crucially on $g_A$
being different from unity \cite{bc79}. 

The purpose of this study is to use the extended linear sigma model 
to derive some of the  
low-energy theorems  for the elastic $\pi$N scattering amplitude, 
to calculate the $\pi$N scattering lengths, and to discuss the 
nucleon sigma term $\Sigma_N$. 
We believe that at least some of the generally valid predictions 
of chiral symmetry are most economically obtained in this model.
Throughout this paper we shall use the tree approximation,  
save for one illustrative example done at the one-loop self-consistent 
approximation level, shown in Appendix A.  
In order to explore the various possibilities, and to facilitate comparison 
with earlier studies of the Gell-Mann--Levy linear sigma model we
introduce three different chiral symmetry breaking
[$\chi$SB] terms, as in Refs.\cite{camp79,bc79}. For  
two of the three $\chi$SB terms, the effects on the pion's mass 
appear first at the tree level, whereas the third $\chi$SB term's effect
is only visible at the one-loop level, see Appendix A. 

This paper falls into six sections.  
In Section~II we define the extended linear sigma model,  
present the $\chi$SB terms and the canonical field variables, and 
show that the Noether charges close
the chiral algebra although $g_A \ne 1$.
Section~III is devoted to a derivation of the elastic $\pi$N scattering 
amplitude, the Adler consistency condition and 
the scattering lengths.  
In Section~IV we examine the nucleon sigma term $\Sigma_N$, first  
from the (formal) operator point of view 
and second as extracted from the elastic $\pi$N scattering 
amplitude in the first Born approximation to draw 
conclusions from the comparison of the two methods. 
In Section~V we examine the connection with the
effective pion-nucleon chiral perturbation theory, and 
Section~VI  summarizes the results.

\section{The extended linear sigma model}

The extended sigma model is the linear sigma model  
modified by adding a pseudovector pion-nucleon coupling
to the pseudoscalar one \cite{ax96}. This model allows
a nucleon axial current with arbitrary $g_A (\neq 1)$. 
The Lagrangian density of this model is given by
\begin{eqnarray}
{\cal L} &=& \bar{\psi}{\rm i}{\partial{\mkern-10mu}{/}}\psi -
g_{0}\bar{\psi}\left[\sigma + 
{\rm i}\gamma_{5} \bbox{\pi} \cdot \bbox{\tau} \right]\psi  
\nonumber \\
&+& {1 \over 2} \left[\left(\partial_{\mu} \sigma \right)^{2}
+ \left(\partial_{\mu} \bbox{\pi}\right)^{2} \right] 
+ {1 \over 2} \mu_{0}^{2}(\sigma^{2} + \bbox{\pi}^{2}) 
- {\lambda_{0} \over 4}
\left(\sigma^{2} + \bbox{\pi}^{2}\right)^{2} + {\cal L}_{\chi SB} 
\nonumber \\
&+& \left({g_{A} - 1 \over{f_{\pi}^{2}}}\right) \left[
\left(\bar{\psi}\gamma_{\mu} {\bbox{\tau}\over 2}\psi \right) \cdot
\left(\bbox{\pi} \times \partial^{\mu} \bbox{\pi}\right)
+ \left(\bar{\psi}\gamma_{\mu} \gamma_{5} {\bbox{\tau}\over 2}\psi 
\right) \cdot
\left(\sigma \partial^{\mu} 
\bbox{\pi} - \bbox{\pi} \partial^{\mu} \sigma \right) \right] ~.
\label{e:lag3} \
\end{eqnarray}
We assume that the parameters 
$\lambda_{0}$ and $\mu_{0}^{2}$ are positive, which 
ensures spontaneous symmetry breaking in the 
tree approximation level.
The last line in Eq.(\ref{e:lag3})
is a non-renormalizable derivative-coupling term, 
introduced by Bjorken and Nauenberg and by Lee\cite{bn68}.  
We shall focus on some consequences of adding this term to the linear
sigma model lagrangian. 

The chiral symmetry breaking ($\chi$SB) 
terms in the Lagrangian are for illustration 
those discussed in Refs.\cite{camp79,bc79}
\begin{equation}
{\cal L}_{\chi SB} = - {\cal H}_{\chi SB} = \varepsilon_{1} \sigma - 
\varepsilon_{2} \bbox{\pi}^2  - \varepsilon_{3} {\bar \psi} \psi~. 
\label{e:chisb}
\end{equation}
An example of a different $\chi$SB term is discussed in, 
e.g.  Ref.\cite{delor96}.
Each one of the three terms in Eq.(\ref{e:chisb}) 
separately breaks the chiral symmetry and is 
capable of shifting the pion mass, though not always in the tree approximation. 
Yet the  three terms do not always predict the same physics in all specific 
instances. In particular they predict different shifts of the nucleon mass,
see Ref. \cite{camp79}, and, e.g. we find a  different 
Goldberger-Treiman [GT] relations: 
$g_{A} M = g_{\pi N} f_\pi + \varepsilon_{3}$. 
%{\bf Is the sign correct? Now I find +}.

As usual we choose the ground state of the model as the  
minimum of the meson interaction lagrangian  
${\cal L}_{\rm meson}^{\rm int}$ w.r.t. the $\sigma-$ and $\pi-$ fields. 
This means shifting the sigma field by its vacuum expectation value, 
$\langle \sigma \rangle_0 \equiv f_\pi$, i.e. $\sigma$ = $f_\pi$ + $s$,
where from the minimum requirement obtain 
\begin{eqnarray}
\left(\mu_{0}^2 - \lambda_0 f_{\pi}^2 \right) f_{\pi} &=& - \varepsilon_{1}~.
\label{e:min} \
\end{eqnarray}
The meson interaction lagrangian in the new field variable reads
\begin{eqnarray}
- {\cal L}_{\rm meson}^{\rm int} &=& 
{1 \over 2} \left(m_{\sigma}^{2} s^{2} + m_{\pi}^2 \bbox{\pi}^{2}\right) + 
{1 \over{2 f_{\pi}}} \left(m_{\sigma}^{2} - m_{\pi}^2 + 2 \varepsilon_{2}\right)
s \left(s^2 + \bbox{\pi}^{2}\right) \nonumber \\
&+&  
{1 \over{8 f_{\pi}^2 }}
\left(m_{\sigma}^{2} - m_{\pi}^2 + 2 \varepsilon_{2} \right) 
\left(s^2 + \bbox{\pi}^{2}\right)^{2}. \
\label{e:potnc}
\end{eqnarray}
The resulting nucleon, $s$-meson and pion 
%($\bbox{\pi}$) 
masses are
\begin{mathletters}
\begin{eqnarray}
M &=& \varepsilon_{3} + g_0 f_{\pi} 
\label{e:na}\\
m_{\sigma}^2 &=&  - \mu_{0}^2 + 3 \lambda_0 f_{\pi}^2 
\label{e:nb}\\
m_{\pi}^2 &=&  - \mu_{0}^2 + \lambda_0 f_{\pi}^2 + 2 \varepsilon_{2}
= \varepsilon_1/f_\pi + 2 \varepsilon_2~. \
\label{e:nc} \
\end{eqnarray}
\end{mathletters} 
The axial-vector Noether current 
\begin{eqnarray}
{\bf J}_{\mu 5}^{a} &=& 
\left(\bar{\psi}\gamma_{\mu} \gamma_{5} {\bbox{\tau}\over 2}\psi 
\right)^{a} -
\left(\bbox{\pi} \partial_{\mu} \sigma - \sigma \partial_{\mu} 
\bbox{\pi}\right)^{a}
\nonumber \\
&+& 
\left({g_{A} - 1 \over{f_{\pi}^{2}}}\right) \Bigg[
\left(\bar{\psi}\gamma_{\mu} \gamma_{5} 
{\bbox{\tau}\over 2}\psi 
\cdot \bbox{\pi} \right) \bbox{\pi}^{a} 
\nonumber \\
&+& 
\sigma^{2}
\left(\bar{\psi}\gamma_{\mu} \gamma_{5} 
{\bbox{\tau}\over 2}\psi \right)^{a} 
+ \sigma
\left(\bar{\psi} \gamma_{\mu} {\bbox{\tau}\over 2}\psi 
\times \bbox{\pi} \right)^{a} \Bigg]  ~,
\label{e:axi4} \
\end{eqnarray}
is partially conserved in this model. The divergence of this axial current is 
\begin{eqnarray}
\partial^{\mu} {\bf J}_{\mu 5}^{a} &=& 
\left(\varepsilon_{1} + 2 \varepsilon_{2} \sigma \right)\bbox{\pi}^{a}  
- \varepsilon_{3} \bar{\psi} i \gamma_{5}\bbox{\tau}^{a}\psi  ~.
\label{e:divax} \
\end{eqnarray}
When we assume that the physical one-pion state, $| \bbox{\pi} \rangle$, 
does not have any $|s \; \bbox{\pi} \rangle$  
or $| N {\bar N} \rangle$ components, the matrix element of the 
divergence of the axial current (using $\sigma = f_{\pi} + s$) 
for the one-pion-to-vacuum transition gives 
\begin{eqnarray}
m_{\pi}^2 f_{\pi} &=& \varepsilon_{1} + 2 f_{\pi}
\varepsilon_{2}~, \
\label{e:nc1} \
\end{eqnarray}
To see explicitly that 
the purely one-nucleon part of the axial current has acquired the
coupling constant $g_{A} \neq 1$, 
Eq. (\ref{e:axi4}) is rewritten with the 
shifted sigma field ($\sigma = f_{\pi} + s$), and we obtain:  
\begin{eqnarray}
{\bf J}_{\mu 5}^{a} &=& g_{A}
\left(\bar{\psi}\gamma_{\mu} \gamma_{5} {\bbox{\tau}\over 2}\psi 
\right)^{a} 
+ f_{\pi} \partial_{\mu} \bbox{\pi}^{a}
+ \left(s \partial_{\mu} \bbox{\pi} 
- \bbox{\pi} \partial_{\mu} s \right)^{a} 
\nonumber \\
&+& 
\left({g_{A} - 1 \over{f_{\pi}^{2}}}\right) \Bigg[
\left(\bar{\psi}\gamma_{\mu} \gamma_{5} 
{\bbox{\tau}\over 2}\psi 
\cdot \bbox{\pi} \right) \bbox{\pi}^{a} 
\nonumber \\
&+&  
s \left(2 f_{\pi} + s \right)
\left(\bar{\psi}\gamma_{\mu} \gamma_{5} 
{\bbox{\tau}\over 2}\psi \right)^{a} 
+ \left(f_{\pi} + s \right)
\left(\bar{\psi} \gamma_{\mu} {\bbox{\tau}\over 2}\psi 
\times \bbox{\pi} \right)^{a} 
\Bigg]  ~.
\label{e:axi5} \
\end{eqnarray}
The axial charge density, however, 
\begin{eqnarray}
\rho_{5}^{a} &=& {\bf J}_{0 5}^{a} =
\psi^{\dagger} \gamma_{5} {\bbox{\tau}^{a}\over 2}\psi 
- \left(\bbox{\pi}^{a} \Pi_{\sigma} - 
\sigma \bbox{\Pi}_{\pi}^{a} \right) ~,
\label{e:axich} \
\end{eqnarray}
retains its linear sigma model form when written in terms of 
canonical fields and their associated canonical momenta \cite{sato98}:
\begin{mathletters}
\begin{eqnarray}
\Pi_{\sigma}  &=& \dot{\sigma} - 
\left({g_{A} - 1 \over{f_{\pi}^{2}}}\right) 
\left(\psi^{\dagger} \gamma_{5}
{\bbox{\tau} \cdot \bbox{\pi}\over 2}\psi \right) 
\\
\Pi_{\pi}^{a} &=& \dot{\bbox{\pi}^{a}} + 
\left({g_{A} - 1 \over{f_{\pi}^{2}}}\right) 
\left[\left(\psi^{\dagger}
{\bbox{\tau} \times \bbox{\pi}\over 2}\psi \right)^{a} + 
\sigma \psi^{\dagger} \gamma_{5} 
{\bbox{\tau}^{a}\over 2}\psi \right]  ~.
\label{e:ax8} \
\end{eqnarray}
\end{mathletters}
The axial charge density Eq.(\ref{e:axich}) 
and the vector  charge density
\begin{eqnarray}
\rho^{a} &=& {\bf J}_{0}^{a} =
\psi^{\dagger} {\bbox{\tau}^{a}\over 2}\psi 
+ \varepsilon^{abc} \left(\bbox{\pi}^{b} \bbox{\Pi}_{\pi}^{c} \right) ~,
\label{e:isoch} \
\end{eqnarray}
close the algebra: 
\begin{mathletters}
\begin{eqnarray}
\left[ \rho^{a}(0, {\bf x}), \rho^{b}(0, {\bf y}) \right] 
&=& i \varepsilon^{abc} \rho^{c}(0, {\bf x}) \delta \left({\bf x} - {\bf y}
\right)~,
\label{e:ax10a}
%\nonumber 
\\
\left[ \rho_{5}^{a}(0, {\bf x}), \rho_{5}^{b}(0, {\bf y}) \right] 
&=& 
i \varepsilon^{abc}\rho^{c}(0, {\bf x}) \delta \left({\bf x} - {\bf y}\right)
~,
\label{e:ax10b}
%\nonumber 
\\
\left[ \rho_{5}^{a}(0, {\bf x}), \rho^{b}(0, {\bf y}) \right] 
&=& 
i \varepsilon^{abc}\rho_{5}^{c}(0, {\bf x}) 
\delta \left({\bf x} - {\bf y}\right) ~,
\label{e:ax10c} \
\end{eqnarray}
\end{mathletters}
when we assume the canonical anti-commutation relations: 
\begin{mathletters}
\begin{eqnarray}
\left\{ \psi^{a}(0, {\bf x}), \Pi_{\bbox{\psi}}^{b}(0, {\bf y}) \right\}  
 &=& 
%\left[ \bbox{\pi}^{a}(0, {\bf x}), \dot{\bbox{\pi}}^{b}(0, {\bf y}) \right] = 
i \delta^{ab} \delta \left({\bf x} - {\bf y}\right) ~.
\label{e:ax9a} \\
\left[ \sigma (0, {\bf x}), \Pi_{\sigma}(0, {\bf y}) \right] 
&=& 
i \delta \left({\bf x} - {\bf y}\right)
\label{e:ax9b} \\
\left[ \pi^{a}(0, {\bf x}), \Pi_{\bbox{\pi}}^{b}(0, {\bf y}) 
\right]  
 &=& 
%\left[ \bbox{\pi}^{a}(0, {\bf x}), \dot{\bbox{\pi}}^{b}(0, {\bf y}) \right] = 
i \delta^{ab} \delta \left({\bf x} - {\bf y}\right) ~.
\label{e:ax9c} \
\end{eqnarray}
\end{mathletters}
Thus we see that in this extended sigma model 
only the spatial part of the nucleon axial
current is renormalized and  
the algebra of the charge operators 
is satisfied.

\section{The elastic $\pi$N scattering amplitude}

We  follow the discussion and methods of 
the linear sigma model in Ref.\cite{camp79}, but extended 
to include the  new terms in the Lagrangian 
shown in the last line of Eq.(\ref{e:lag3}).
The main consequence of this modified Lagrangian is that the original $\pi$N
coupling constant $g_0$ is renormalized to 
$g_{\pi N} = g_0 \left[1 + 
\left(g_{A} - 1 \right)\left({M \over{g_{0} f_{\pi}}}\right) \right]$, where
the nucleon mass is  $M = g_{0} f_\pi + \varepsilon_{3}$.
This leads to a different set of S-wave scattering lengths and to a 
change in the GT relation written above.
Otherwise in the tree approximation the nucleon $\Sigma$ terms 
are identical to those found by Campbell\cite{camp79} as we show below. 

\subsection{The scattering amplitude}

The elastic $\pi$N scattering amplitude $T$ is usually written 
in terms of its two
isospin and two Dirac matrix components as follows
\begin{eqnarray}
T_{\alpha \beta} &=& T^{(+)} \delta_{\alpha \beta} + T^{(-)}
{1 \over 2} \left[\bbox{\tau}_{\alpha}, \bbox{\tau}_{\beta}\right]
\nonumber \\ 
T &=& A + B{1 \over 2}
\left( {k{\mkern-10mu}{/}}_{1} + {k{\mkern-10mu}{/}}_{2}\right)~, \
\label{e:tmatrix1}
\end{eqnarray}
where the incoming and outgoing pion's momenta are $k_1$ and $k_2$, and $\alpha$
and $\beta$ are their isospin indices.
%(assuming all the initial and final state particles to be on their mass-shells). 
An explicit calculation of the four tree-level 
diagrams in Fig. \ref{fig1} leads to
\begin{mathletters}
\begin{eqnarray}
A^{(+)} &=& \left({g_{0} \over{f_{\pi}}}\right)
\left[\left({m_{\sigma}^2 - m_{\pi}^2 + 2 \varepsilon_{2} 
\over{m_{\sigma}^2 - t}}\right) 
+ 2 \left(g_{A} - 1 \right)
%\left({M \over{g_{0} f_{\pi}}}\right) 
+ \left(g_{A} - 1 \right)^{2}\left({M \over{g_{0} f_{\pi}}}\right) \right]
\label{e:tmatrix2a}
%\nonumber 
\\ 
A^{(-)} &=& 0
\label{e:tmatrix2b}
%\nonumber 
\\ 
B^{(+)} &=& 
g_{0}^2 \left[1 + \left(g_{A} - 1 \right)\left({M \over{g_{0} f_{\pi}}}\right) 
\right]^2
\left[{1 \over{M^2 - s}} - {1 \over{M^2 - u}}\right]
\label{e:tmatrix2c}
%\nonumber 
\\ 
B^{(-)} &=& 
%g_{\pi N}^2 
g_{0}^2 \left[1 + 
\left(g_{A} - 1 \right)\left({M \over{g_{0} f_{\pi}}}\right) \right]^2
\left[{1 \over{M^2 - s}} + {1 \over{M^2 - u}} \right] - 
{1 \over{2 f_{\pi}^{2}}} \left(g_{A}^{2} - 1 \right) ~, \
\label{e:tmatrix2d}
\end{eqnarray}
\end{mathletters}
where $s$, $t$ and $u$ are the standard Mandelstam variables, and 
$s + t + u = 2 M^2 + k_1^2 + k_2^2$.
Below we will use the traditional kinematical 
variables in the expressions for the amplitudes: 
\begin{mathletters}
\begin{eqnarray}
\nu &=& {1\over{4M}}\left(k_{1} + k_{2}\right) \cdot \left(p_{1} + p_{2}\right)
%\nonumber \\ &=& 
= {s - u \over{4M}}
\label{e:nu}\\
\nu_B &=& 
- {k_{1} \cdot k_{2} \over{2 M}}
%\right)\nonumber \\ &=& 
= {t - k_{1}^2 - k_{2}^2 \over{4 M}}~. \
\label{e:nub}
\end{eqnarray}
\end{mathletters}
We follow standardnotation and use $ D^{(\pm)}$ as an abbreviation for 
\begin{eqnarray}
D^{(\pm)} &\equiv& A^{(\pm)} + \nu B^{(\pm)}~. \
\label{e:dampl1}
\end{eqnarray}
In the tree approximation
the extended sigma model isospin antisymmetric amplitude $D^{(-)}$  then 
reads for on- and off-mass-shell pions 
\begin{eqnarray}
D^{(-)}(\nu, \nu_{B}, k_{1}^2, k_{2}^2 ) 
&=& 
\left({g_{\pi N}^2 \over{M}}\right) 
\nu \left[{\nu_{B} \over{\nu_{B}^{2} - \nu^{2}}} - 
\left({1 - g_{A}^{-2} \over{2 M }}\right) 
\left(1+g_A^{-1} \frac{2\varepsilon_3}{M}\right) \right] 
+ {\cal O}(\nu \varepsilon_i^2) 
%\nonumber \\
%D_{\rm PV Born}^{(-)}(\nu, \nu_{B}, k_{1}^2, k_{2}^2)
%&=& 
%\left({g_{\pi N}^2 \over{M}}\right) 
%\left({\nu \nu_{B} \over{\nu_{B}^{2} - \nu^{2}}} \right) 
%+ {\cal O}(\varepsilon_i^2)
~, \
\label{e:d-}
\end{eqnarray}
where we have used our GT relation. The second term in the square bracket 
$\propto (1-g_A^{-2})/2M$ is absent in the regular sigma models where $g_A =1$. 
%the subscript ``PV Born" denotes the Born amplitude obtained by 
%using a purely pseudovector (PV) $\pi N$ interaction lagrangian.
To obtain the tree-level isospin symmetric amplitude we rewrite 
Eq. (\ref{e:tmatrix2a}) as follows 
\begin{eqnarray}
A^{(+)} &=& 
\left({g_{\pi N}^2 \over{M}}\right) 
\left[1 -  g_{A}^{-2}
\left({m_{\pi}^2 - t - 2 \varepsilon_{2}\over{m_{\sigma}^2 - t}}\right)
+  g_{A}^{-2} {\varepsilon_{3} \over{M}}
%+  g_{A}^{-2} {\varepsilon_{3} \over{M}} (1-\frac{2}{g_A} )  
%\left(\frac{m_\pi^2 -t}{m_\sigma^2 -t}\right)
 \right] + {\cal O}(\varepsilon^{2})~,
\label{e:a+2} \
\end{eqnarray}
%where for $t$ = ${\cal O}(m_\pi^2)$ the last term is ${\cal O}(\varepsilon )$
%since $m_\pi^2 \sim \varepsilon_i$, i = 1, 2, {\bf Eq.(6c)???}.
and Eq. (\ref{e:tmatrix2c}) is rewritten as 
\begin{eqnarray}
B^{(+)} &=& 
{g_{\pi N}^2 \over{M}} 
{\nu \over{\nu_{B}^{2} - \nu^{2}}} 
+ {\cal O}(\varepsilon^{2})~,
\label{e:b+2} \
\end{eqnarray}
which gives, even for off-mass-shell pions: 
\begin{eqnarray}
D^{(+)}(\nu, \nu_{B}, k_{1}^2, k_{2}^2) 
&=& 
\left({g_{\pi N}^2 \over{M}}\right) 
\Bigg\{{\nu_{B}^{2} \over{\nu_{B}^{2} - \nu^{2}}} - g_{A}^{-2}
\left({m_{\pi}^2 - t - 2 \varepsilon_{2}\over{m_{\sigma}^2 - t}}\right) 
%\nonumber \\ & & 
%+ g_{A}^{-2} {\varepsilon_3 \over{M}}\left(1+ 
+ g_A^{-2}\frac{\varepsilon_3}{M} 
%(1-\frac{2}{g_A}) 
%\left(\frac{m_\pi^2 -t}{m_\sigma^2 -t} \right) \right)
\Bigg\} + {\cal O}(\varepsilon^{2}) ~. \
\label{e:st}
\end{eqnarray}
Note that Eq.(\ref{e:st}) is zero for $\nu = \nu_B =0$ and $t = m_\pi^2$ 
only if $\varepsilon_2 = \varepsilon_3$ = 0. 
This means both 
the isospin symmetric $D^{(+)}$ and  antisymmetric  $D^{(-)}$  amplitudes
have Adler zeros 
{\it only if PCAC, in its narrow definition, is satisfied as an 
operator equation} 
in the extended sigma model. 
This can also be seen by following Campbell's analysis\cite{camp79} 
of the original linear sigma model. 
The main difference from the original linear sigma model is that the $\pi$N 
coupling constant is 
renormalized from $g_0$ in the original linear $\sigma$ model to 
$g_{\pi N} = g_{A} g_{0} \left[1 + {\varepsilon_{3} \over{g_{0} f_{\pi}}}
\left(1 - g_{A}^{-1}\right)\right]$, see Eq.(\ref{e:tmatrix2c}), and that  the 
GT relation becomes $g_A M = g_{\pi N} f_\pi + \varepsilon_{3}$, after 
``turning on'' $\varepsilon_{3} \ne 0$, i.e., at the tree-level the 
GT relation acquires an ``anomaly'' $\propto \varepsilon_3$.  
%{\bf Should it beopposite sign:$g_A M = g_{\pi N} f_\pi +\varepsilon_{3}$ ??}

\subsection{Scattering lengths}

The $\pi N$ scattering lengths are given by Eqs.(\ref{e:tmatrix2a}-d): 
\begin{eqnarray}
a_{0}^{(\pm)} &=& 
{D_{\rm threshold}^{(\pm)} \over{4 \pi \left(1 + {m_{\pi} \over M}\right)}}
=
{1 \over{4 \pi \left(1 + {m_{\pi} \over M}\right)}}
\left[ A^{(\pm)} + m_{\pi} B^{(\pm)}\right]_{\rm threshold}  ~, \
\label{e:scatl}
\end{eqnarray}
which leads to the standard result for the isospin-symmetric scattering length 
in the $\varepsilon_{2} = \varepsilon_{3} = 0$
(but $\varepsilon_{1} \ne 0$ since $m_\pi \ne 0$) limit
\begin{eqnarray}
a_{0}^{(+)} &=& 
{g_{\pi N}^2 \over{4 \pi \left(1 + {m_{\pi} \over M}\right)}}
\left({m_{\pi} \over M}\right)
\left[1 - {1 \over{1 - \left({m_{\pi} \over2 M}\right)^{2}}}
- g_{A}^{-2} \left({m_{\pi} \over m_{\sigma}}\right)^2
\right]{1 \over{m_{\pi}}} 
\nonumber \\
&\simeq& 
{g_{\pi N}^2 \over{4 \pi \left(1 + {m_{\pi} \over M}\right)}}
\left({m_{\pi} \over M}\right)
\left[ - \left({m_{\pi} \over2 M}\right)^{2}
- g_{A}^{-2} \left({m_{\pi} \over m_{\sigma}}\right)^2
\right]{1 \over{m_{\pi}}} ~. \
\label{e:scatl+}
\end{eqnarray}
The value for $a_{0}^{(+)}$ is smaller than the 
value in the ordinary linear $\sigma$ model, 
see e.g. Delorme {\it et al.} \cite{delor96}, due to  
the factor $g_{A}^{-2}$ 
in front of the second term.  This will be discussed further in section IV C.

The isospin antisymmetric scattering length equals  
the standard Weinberg-Tomozawa result
\begin{eqnarray}
a_{0}^{(-)} &=& 
{g_{\pi N}^2 \over{4 \pi \left(1 + {m_{\pi} \over M}\right)}}
\left({m_{\pi} \over M}\right)^2
\left[{1 \over{1 - \left({m_{\pi} \over{2 M}}\right)^{2}}}
- \left(1 - g_{A}^{-2} \right)\right]
\left({1 \over{2 m_{\pi}}}\right) + {\cal O}(\varepsilon_3^2) 
\nonumber \\
&\simeq& 
{g_{\pi N}^2 \over{8 \pi \left(1 + {m_{\pi} \over M}\right)}}
\left({m_{\pi} \over M}\right)^2
\left({1 \over{g_{A}^{2} m_{\pi}}}\right) + \cdots
%{\cal O}(\varepsilon_3^2) 
~. \
\label{e:scatl-}
\end{eqnarray}

In the  case when $\varepsilon_{i} \ne 0$, $i$ = 1,2,3, we have 
\begin{eqnarray}
a_{0}^{(+)} 
&\simeq& 
{- g_{\pi N}^2 \over{4 \pi \left(1 + {m_{\pi} \over M}\right)}}
\left({m_{\pi} \over M}\right)
\Bigg\{\left({m_{\pi} \over2 M}\right)^{2} 
\nonumber \\ &+&  
 g_{A}^{-2} 
\left[\left({m_{\pi}^2 - 2 \varepsilon_{2} \over m_{\sigma}^2 }\right)
- \left({ \varepsilon_{3} \over M}\right) 
%\left(1+ (1-\frac{2}{g_A}) \frac{m_\pi^2}{m_\sigma^2}\right)
 \right] \Bigg\}{1 \over{m_{\pi}}} 
+ {\cal O}(\varepsilon_i^2)~, \
\label{e:scatl+1}
\end{eqnarray}
for the isoscalar scattering length.  
%where again the last term $\propto (1-2/g_A)$ is ${\cal O}(\varepsilon^2)$.
Note the negative sign in front of the $\varepsilon_3$ term which allows for
either sign of this scattering length. The isovector scattering length is 
%when $\varepsilon_{3} \ne 0$  
\begin{eqnarray}
a_{0}^{(-)} &\simeq& 
{g_{\pi N}^2 \over{8 \pi \left(1 + {m_{\pi} \over M}\right)}}
\left({m_{\pi} \over M}\right)^2  \; {1 \over{g_{A}^{2}}}
%\left[{1 \over{g_{A}^{2}}} - 2 {\varepsilon_{3} \over M} g_A^{-1} 
%\left(1 - g_{A}^{-2}\right) \right] 
{1 \over{m_{\pi}}} + {\cal O}(\varepsilon_3^2) ~, \
\label{e:scatl-1}
\end{eqnarray}
unchanged from the Weinberg-Tomozawa result. 
%{\bf where again the last term $\propto (1-g_A^{-2})$ is 
%${\cal O}(\varepsilon^2)$.}
%
To compare these results with experiment
we will determine the values of the $\chi$SB coefficients 
$\varepsilon_{i}$
from some other source,  see Sect. IV.C and appendix B. 
However, as all three $\chi$SB terms in Eq.(\ref{e:chisb}) with their 
full strengths are not possible without overcounting, some care 
with the interpretation of these results is necessary. 

\section{The sigma term}

The pion-nucleon $\Sigma_{N}$ term is of importance 
for investigations of the $\langle \bar{\psi} \psi \rangle$ 
condensate in nuclear matter\cite{vd99,delor96},  
and in determination of the flavour content of the 
nucleon \cite{kj87}. 
As we shall show the extended sigma model gives a very 
interesting answer to the question of the flavour content
of the nucleon.
First we discuss the nucleon $\Sigma_N$ term as obtained from the $\Sigma$ 
operator and then evaluate the nucleon $\Sigma_N$ term from
the $\pi N$ amplitude. 
We shall  show  that 
the connected one-nucleon matrix element 
of the $\Sigma$ term operator coincides with
another (operational) definition of the nucleon
$\Sigma$ term based on the pion-nucleon elastic scattering amplitude
in the tree approximation. 
Finally we make a short estimate of the 
possible values of the $\Sigma_N$ term in this model 
and also discuss the possible values of the $\pi N$ scattering
lenghts. 

\subsection{Operator definition}

The $\Sigma$ operator is defined as 
\begin{eqnarray}
\Sigma^{ab} &=& 
\big[Q^{a}_{5}, [Q^{b}_{5}, {\cal H}_{\chi SB}]\big] 
\nonumber \\ 
\Sigma &=& 
{1 \over 3} \sum_{a=b=1}^{3} \Sigma^{ab}  ~. \
\label{e:sigma1}
\end{eqnarray}
Using  a chiral Ward identity  this operator  
appears after two applications of Sakurai's ``master formula" 
to any elastic S-matrix element with one pion in the initial 
and one in the final state \cite{sak69,dash69}. 
Here $a, b$ are the flavour indices of the axial charge
$Q_{5}^{a} = \int d{\bf x} \rho_{5}^a$
appropriate to the corresponding pseudoscalar mesons (pions), and 
${\cal H}_{\chi SB}$ is the chiral symmetry-breaking Hamiltonian density.
In principle all of the
objects entering Eq. (\ref{e:sigma1}) are meant to be exact Heisenberg
representation operators. 
%(or states when matrix elements are involved). 
As we do not have exact solutions 
to the quantum-field equations of motion, we 
%must often content ourselves with
will discuss two approximate 
%solutions to the 
matrix elements of the $\Sigma^{ab}$-operator for  
two cases: (i) the vacuum expectation value 
$\langle 0| \Sigma |0 \rangle$, and (ii)
the nucleon expectation value of its volume integral
$\langle N| \int d{\bf x} \Sigma(x) |N \rangle$. 
The vacuum matrix element is well understood \cite{reya74}, 
so it leads to valuable constraints on the form of the 
$\chi$SB terms.  
%insights that can only be obtained from this source. 
As for the nucleon matrix element, we compare the results
obtained from the above operator definition using the 
equations-of-motion, with another derivation based on the 
off-shell elastic $\pi N$ scattering amplitude.

\subsubsection{The $\Sigma$ vacuum expectation value}

The vacuum expectation value of the $\Sigma$ operator 
yields Dashen's formula \cite{dash69}
\begin{equation} 
\left(f m^{2} f \right)^{ab} = f_{a} m^{2}_{ab} f_{b} 
= - \langle 0 \left| 
\big[Q^{a}_{5}, [Q^{b}_{5}, {\cal H}_{\chi SB}]
\big] \right| 0 \rangle ~. \
\label{e:dash1}
\end{equation}
This formula describes the lowest order $\chi$SB 
correction to the otherwise vanishing pseudoscalar meson mass squared 
($m_{ps}^2$) for arbitrary chiral symmetry-breaking terms 
in the  Hamiltonian density ${\cal H}_{\chi SB}$. 
When the $\chi$SB term is taken
to be the current quark mass in the QCD hamiltonian
$${\cal H}_{q\chi SB} = {\bar q} m_{q}^{0}q =
m_{u}^{0} {\bar u}u + m_{d}^{0} {\bar d}d , $$ 
Eq.(\ref{e:dash1}) yields 
\begin{eqnarray}
\left(f_{\rm ps} m_{\rm ps}^{2} f_{\rm ps} \right)^{ab} 
&=& 
%- \langle 0 \left| \Sigma^{ab} \right| 0 \rangle 
%\nonumber \\ &=& 
- \langle 0 \left| \left[Q^{a}_{5}, \left[Q^{b}_{5}, 
{\bar q} m_{q}^{0}q \right]\right] \right| 0 \rangle 
\nonumber \\
&=& 
- \langle 0 |{\bar q}
\left\{ \left\{ m_{q}^{0}, {\lambda^a \over 2}
\right\}, \frac{\lambda^b}{2} \right\}q | 0 \rangle ~, \
\label{e:dash2}
\end{eqnarray}
where $\lambda^a$, are the Gell-Mann matrices. 
By averaging over $a=1, 2, 3$  
one finds the Gell-Mann--Oakes--Renner (GMOR) relation 
between the pion mass and decay constant on one hand and the current 
quark mass Hamiltonian vacuum expectation value on the other: 
%\begin{mathletters} 
\begin{eqnarray} 
m_{\pi}^{2} f_{\pi}^{2} &=& - \left[ 
m_{u}^{0} \langle 0 |\bar{u} u | 0 \rangle  +
m_{d}^{0} \langle 0 |\bar{d} d | 0 \rangle \right] 
\label{e:gmora} ~,\
\end{eqnarray}
%\end{mathletters}

To make contact with our previous discussion  
we apply Eq.(\ref{e:dash1}) to 
our extended sigma model with the
three kinds of  $\chi$SB terms of 
Eq.(\ref{e:chisb}). We  use the canonical commutation relations
Eq.(\ref{e:ax9a},b) and the axial charge Eq.(\ref{e:axich}) to obtain:
\begin{eqnarray}
\big[Q^{a}_{5}, [Q^{b}_{5}, {\cal H}_{\chi SB}(0)]\big] &=&
- \varepsilon_{1} \sigma \delta^{ab} - 
2 \varepsilon_{2} \left(\sigma^2 \delta^{ab} - 
\bbox{\pi}^{a} \bbox{\pi}^{b}\right) + 
\varepsilon_{3} {\bar \psi}\psi \delta^{ab}~. \
\label{e:dash3}
\end{eqnarray}
Taking the vacuum expectation value of this expression we find
\begin{equation} 
\left(m_{\pi} f_{\pi} \right)^{2} = 
\varepsilon_{1} \langle 0 | \sigma | 0 \rangle + 2 \varepsilon_{2}
\langle 0 | \sigma^{2} | 0 \rangle - 
\varepsilon_{3} \langle 0 | {\bar \psi}\psi | 0 \rangle~, 
\label{e:dash4}
\end{equation} 
This relation goes beyond the tree approximation of Eq.(\ref{e:nc}) 
as we show in appendix A. 
We shall first examine Eq.(\ref{e:dash4}) for the three distinct types of the
$\chi$SB Hamiltonian in order to determine/normalize the values of
the coefficients $\varepsilon_{i}$.

(i) $\varepsilon_i = 0$ for i = 2 and 3 
leads to 
\begin{equation}
\varepsilon_{1} \langle 0  \left| \sigma \right| 0 \rangle = 
\left(m_{\pi} f_{\pi} \right)^{2}~, 
\end{equation}
i.e., $\varepsilon_{1} = m_{\pi}^{2} f_{\pi}$. 

(ii) $\varepsilon_i = 0$ for i = 1 and 3 
leads to 
\begin{equation}
2 \varepsilon_{2} \langle 0 | \sigma^2 | 0 \rangle = 
\left(m_{\pi} f_{\pi} \right)^{2}~, 
\end{equation}
i.e., $\varepsilon_{2} = {1 \over 2}
m_{\pi}^{2}~. $

(iii) $\varepsilon_i = 0$ for i = 1 and 2 
leads to the relation
\begin{equation}
- \varepsilon_{3} \langle 0 |{\bar \psi}\psi | 0 \rangle = 
\left(m_{\pi} f_{\pi} \right)^{2}~. 
\end{equation} 

We remark that 
this last relation looks like a nucleonic version of the GMOR relations 
Eq.(\ref{e:gmora}). To make this analogy more obvious, we introduce 
the explicit $\chi$SB ``bare" nucleon mass matrix in our  
extended sigma model lagrangian, Eq.(\ref{e:lag3}), and compare it 
with ${\cal L}_{\chi SB}$, Eq.(\ref{e:chisb}). 
The corresponding $\chi$SB Hamiltonian 
density
$${\cal H}_{N\chi SB} = {\bar \psi} M_{N}^{0}\psi =
M_{p}^{0} {\bar p}p + M_{n}^{0} {\bar n}n, $$ 
is used in Eq.(\ref{e:dash1})
to obtain the relation 
\begin{eqnarray} 
m_{\pi}^{2} f_{\pi}^{2} &=& - 
\left[M_{p}^{0} \langle \bar{p} p \rangle_{0} +
M_{n}^{0} \langle \bar{n} n \rangle_{0} \right] ~~. \
\label{e:gmorb} 
\end{eqnarray}
The obvious conclusion is  that 
$\varepsilon_{3} = M_N^{0}$, the averaged 
``bare" nucleon mass, as expected from Eqs.(\ref{e:chisb}) or (\ref{e:na}). 
We naturally express 
%the isospin averaged bare nucleon mass 
%${\bar M}_{N}^{0} = {1 \over 2} \left(M_{p}^{0} + M_{n}^{0}\right)$, 
in terms of the  current quark masses: 
$ M_{N}^{0} = 3 {\bar m}_{q}^{0}$ = ${3 \over 2} 
\left(m_{u}^{0} + m_{d}^{0}\right) \simeq$ 23 MeV.  
%from which the relation 
%$\langle \bar{q} q \rangle_{0} = 3 \langle \bar{N} N \rangle_{0}$
%between the condensates follows. 

The basic underlying assumption of chiral perturbation theory 
as an effective hadronic field theory of QCD is
that the $\chi$SB part of the Hamiltonian is a small perturbation.  
Two theories with different degrees of freedom (d.o.f.), e.g. 
quarks in one and  hadrons in another, can be
viewed as effectively mirroring each other 
provided both satisfy the same chiral symmetry transformations. 
For example, 
in a model with hadronic d.o.f.
\footnote{Note that two sets of $\chi$SB terms may effectively mirror
each other under a ``lower" chiral symmetry like  $SU_{L}(2) \times
SU_{R}(2)$, but be very different under a ``higher" symmetry like
$SU_{L}(3) \times SU_{R}(3)$. For example  
the chiral transformation properties of both the current quark 
${\cal H}_{q\chi SB}$ and the bare nucleon mass term 
${\cal H}_{N\chi SB}$ are those of $(2,{\bar 2}) \oplus ({\bar 2},2)$. 
However, in the $N_{f} = 3$ case
the quarks form an SU(3) triplet, which means that 
their bare mass terms transform as
$(3,{\bar 3}) \oplus ({\bar 3},3)$, whereas the spin 1/2 baryons are part of  
an SU(3) octet, which means that their $\chi$SB terms transform as either
$(8,8)$ or  $(8,1) \oplus (1,8)$ under the chiral
$SU_{L}(3) \times SU_{R}(3)$ group \cite{reya74}.
This group theoretical difference implies different pseudoscalar meson mass 
spectra in these two models of $\chi$SB. Since we
know that the observed pseudoscalar masses 
conform rather well with the current
quark mass model \cite{reya74}, we are forced to conclude that the
baryon-antibaryon contribution to the pseudoscalar mass spectrum is
supressed. This raises the question to what extent
one may apply the baryon current mass model of 
$\chi$SB and $\varepsilon_3 \ne 0$ in the two-flavor sector.}
the $\chi$SB part 
due to the current quark mass term in QCD is effectively
mirrored in a pion mass term (plus possibly other terms with the
same transformation properties).  
Chiral perturbation theory goes one step further 
and includes (to a given chiral order) 
all possible $\chi$SB terms in the Hamiltonian. 
The so-called low energy coefficients multiplying these $\chi$SB terms 
are then fit to
the experimental data, though they could also be modelled in 
%/deduced from 
an underlying quark model \cite{meis98}.

In the following we argue that the cases (i), (ii) and (iii) could be 
interchangeable, at least as far as the non-zero pion mass is concerned.
We wish to establish to what extent this interchangeability of the $\chi$SB
terms actually holds in various approximations. 
[They certainly are not equivalent when it comes to non-vacuum matrix elements
of the $\Sigma$ term, as we shall show below.]
In the tree approximation
the first two terms on the r.h.s. of Eq. (\ref{e:dash4}) are the same as
those in Eq. (\ref{e:nc1}). 
Thus we see that the bare (current) nucleon mass term 
with $\varepsilon_3 \ne 0$ does not lead to a 
massive pion in the tree approximation. 
In Appendix \ref{app} we show how the bare nucleon mass 
$\bar{M}_N^0 = \varepsilon_{3} \ne 0$
produces a non-zero pion mass $m_\pi \ne 0$ 
in agreement with the Dashen formula 
Eq.(\ref{e:dash4}) at the one-nucleon-loop 
self-consistent approximation level. 
One immediate conclusion  is that the 
nominally identical forms of $\chi$SB terms in Eq.(\ref{e:dash4}) do not always  
produce the same kinds of effects at the same level of approximation, even 
if the approximations conserve chiral-symmetry in the chiral limit
$\varepsilon_{i} = 0$, $i$ = 1,2,3.

Another consequence of Eq.(\ref{e:dash4}) is that if one assumes the existence 
of more than one $\chi$SB term, then not all of such terms can have their 
``full'' strengths. Specifically, if one wishes to have more than one $\chi$SB 
term in the Hamiltonian Eq.(\ref{e:chisb}),
the coefficients $\varepsilon_{i}$ must be rescaled.  
The new ``scaling 
coefficients'' $\alpha_{i}$ are defined as 
\begin{mathletters}
\begin{eqnarray}
\varepsilon_{1} &=& \alpha_{1} m_{\pi}^{2} f_{\pi}
\label{e:epsa} \\
\varepsilon_{2} &=& \alpha_{2} {1 \over 2} m_{\pi}^{2} 
\label{e:epsb}\\
\varepsilon_{3} &=& \alpha_{3} \bar{M}_N^0~, 
\label{e:epsc} \
\end{eqnarray}
\end{mathletters}
subject to the condition 
of Eq.(\ref{e:dash4}) that $\sum_{i=1}^{3} \alpha_{i} = 1$. 
Similar problems arise
in other quantities sensitive to $\chi$SB terms, 
such as the scattering lengths, 
Eqs.(\ref{e:scatl+1}) and (\ref{e:scatl-1}) as in,  
e.g. Ref.\cite{delor96}.

\subsubsection{The nucleon $\Sigma$ term}

The nucleon $\Sigma$ term ($\Sigma_N$) is, by definition, the connected 
elastic one-nucleon matrix element of the spatial (volume) integral
\footnote{This accounts for the different dimensions of the vacuum- and 
nucleon $\Sigma$ terms. } 
of the $\Sigma$ operator
\begin{eqnarray}
\Sigma_N &=&
\langle N \left| \int d{\bf x} \Sigma (x)\right| N \rangle_{\rm connected}  = 
\langle \int d{\bf x} \Sigma (x) \rangle_N -
\langle \int d{\bf x} \Sigma (x) \rangle^{\rm disconnected}_N 
\nonumber \\
&=&  
{1 \over 3} \sum_{a=b=1}^{3}
\langle N \left| \left[Q^{a}_{5}, \left[Q^{b}_{5}, H_{\chi SB} \right] \right] 
\right| N \rangle 
\nonumber \\
&-&   
(2 \pi)^3 \delta^{(3)}(0) {1 \over 3} \sum_{a=b=1}^{3} \langle 0 \left| 
\left[Q^{a}_{5}, \left[Q^{b}_{5}, {\cal H}_{\chi SB}(0) \right] \right] 
\right| 0 \rangle 
\nonumber \\
&=&  
\int d{\bf x}\Bigg\{\langle \Sigma (x) \rangle_{N} -
\langle \Sigma (0) \rangle_{0}\Bigg\} ~, \
\label{e:dash5}
\end{eqnarray}
where $H_{\chi SB} = \int d{\bf x}{\cal H}_{\chi SB}(x)$. 
%<--- NECESSARY???} 
In this application it is preferable to quantize the system in a finite volume 
$\Omega$, so as to avoid dealing with a new infinity in the form of a Dirac
delta function of zero argument, 
$(2 \pi)^3 \delta^{(3)}(0) = 
\lim_{\Omega \to \infty} (\Omega = \int_{\Omega} d{\bf x})$. 
Subtraction of the disconnected term proceeds naturally 
using the equations of motion. 

Initially we have 
\begin{eqnarray}
{1 \over 3} \sum_{a=b=1}^{3} \int d{\bf x} 
\big[Q^{a}_{5}, [Q^{b}_{5}, {\cal H}_{\chi SB}(x)]\big] 
&=&
\int d{\bf x} \left( - \varepsilon_{1} \sigma - 2 \varepsilon_{2}  
\left(\sigma^2 - {1 \over 3} \bbox{\pi}^{2}\right) 
+ \varepsilon_{3} {\bar \psi}\psi \right)
\nonumber \\
&=&
- \Omega 
\left(\varepsilon_{1} + 2 \varepsilon_{2} f_{\pi}\right) f_{\pi}
- \int d{\bf x} \left(s 
\left(\varepsilon_{1} + 4 \varepsilon_{2} f_{\pi}\right)
- \varepsilon_{3} {\bar \psi}\psi \right)
\nonumber \\
&+&
{\cal O}(s^2) + {\cal O}(\bbox{\pi}^{2}) \; . \
\label{e:sign}
\end{eqnarray}
Using Eq.(\ref{e:potnc}) we obtain 
the equations of motion for the shifted $\sigma$ field $s$:
\begin{eqnarray}
\left[\partial^{2} + m_{\sigma}^2 \right] s 
&=& 
- g_{0} \bar{\psi}\psi -
% \varepsilon_{1} -
\left({m_{\sigma}^{2} - m_{\pi}^2 + 2 \varepsilon_{2} \over{2 f_{\pi}}}\right)
\left(3 s^2 + \bbox{\pi}^{2}\right) 
\nonumber \\
&+& 
\left({m_{\sigma}^{2} - m_{\pi}^2 + 
2 \varepsilon_{2} \over{2 f_{\pi}^2 }}\right) 
s \left(s^2 + \bbox{\pi}^{2}\right)
\nonumber \\
&+& 
\left({g_{A} - 1 \over{2 f_{\pi}^{2}}}\right) \left[ 
\partial^{\mu} \left(\bar{\psi}\gamma_{\mu} \gamma_{5} \bbox{\tau} \cdot
\bbox{\pi}\psi \right) + 
\left(\bar{\psi}\gamma_{\mu} \gamma_{5} \bbox{\tau} \psi 
\right) \cdot
\left(\partial^{\mu} \bbox{\pi} \right) \right] ~.
\label{e:eom1} \
\end{eqnarray}
The lowest-order perturbative solution 
is the following integral equation defined 
in terms of the free Klein-Gordon Green's function 
$\Delta_{F}(x; m_{\sigma})$: 
\begin{eqnarray}
%\left[\partial^{2} + m_{s}^2 \right] 
s(x) &=& 
 \int d^{4}y \Delta_{F}(x - y; m_{\sigma}) \Bigg\{ g_{0}\bar{\psi}(y) \psi(y) +
%- \varepsilon_{1} -
\left({m_{\sigma}^{2} - m_{\pi}^2 + 2 \varepsilon_{2} \over{2 f_{\pi}}}\right)
\left(3 s^2(y) + \bbox{\pi}^{2}(y)\right) 
\nonumber \\
&+& 
\left({m_{\sigma}^{2} - m_{\pi}^2 + 
2 \varepsilon_{2} \over{2 f_{\pi}^2 }}\right) 
s(y) \left(s^2(y) + \bbox{\pi}^{2}(y)\right)
\nonumber \\
&+& 
\left({g_{A} - 1 \over{2 f_{\pi}^{2}}}\right) \left[ 
\partial^{\mu} \left(\bar{\psi}\gamma_{\mu} \gamma_{5} \bbox{\tau} \cdot
\bbox{\pi}\psi \right) + 
\left(\bar{\psi}\gamma_{\mu} \gamma_{5} \bbox{\tau} \psi 
\right) \cdot
\left(\partial^{\mu} \bbox{\pi} \right) \right] \Bigg\}~,
\label{e:eoms2} \
\end{eqnarray}
which upon inserting into the definition (\ref{e:sign}) leads to
\begin{eqnarray}
\Sigma_N &=&
\langle \Sigma \rangle^{\rm connected}_N  =
{1 \over 3} \sum_{a=b=1}^{3} \int d{\bf x} \langle N \left|
\big[Q^{a}_{5}, [Q^{b}_{5}, {\cal H}_{\chi SB}(x)]\big]
\right| N \rangle_{\rm connected}
\nonumber \\
&=&
- \left(\varepsilon_{1} + 4 \varepsilon_{2} f_{\pi}\right) 
\int d{\bf x} \langle s(x) \rangle_{N} 
- \varepsilon_{3} \int d{\bf x} \langle {\bar \psi}(x)\psi(x) \rangle_{N}
+ {\cal O}(s^2) + {\cal O}(\bbox{\pi}^{2})
\nonumber \\
&=&
- g_{0} 
\left(\varepsilon_{1} + 4 \varepsilon_{2} f_{\pi}\right) \int d{\bf x}  
\int d^{4}y \Delta_{F}(x - y; m_{s}) \langle \bar{\psi}(y) \psi(y) \rangle_{N}
\nonumber \\
&+&
\varepsilon_{3} \int d{\bf x} \langle {\bar \psi}(x)\psi(x) \rangle_{N}
+ {\cal O}(s^2) + {\cal O}(\bbox{\pi}^{2})
\nonumber \\
&=&
{g_{0} \over m_{\sigma}^{2}}
\left(\varepsilon_{1} + 4 \varepsilon_{2} f_{\pi}\right) 
+ \varepsilon_{3} + 
{\cal O}(\varepsilon^2) 
\cdots ~.\
\label{e:sign1}
\end{eqnarray}
where the dots represents higher order terms of the fields which are neglected 
since we are working within   the tree aproximation. 
Using Eq.(\ref{e:na}) and 
the tree approximation result Eq.(\ref{e:nc1}), we find
\begin{eqnarray}
\Sigma_N &=&
M \left({m_{\pi}^{2} + 2 \varepsilon_{2} \over m_{\sigma}^{2}}\right) 
+ \varepsilon_{3} + {\cal O}(\varepsilon^2)~.
\label{e:sign2} \
\end{eqnarray}
Naively we expect the value of $\Sigma_N$ to be given by 
the sum of the current quark masses\cite{kj87} $\propto \varepsilon_{3}$ 
which is reflected in a nonzero ``bare" nucleon mass. 
The presence of the scalar field which induces 
the spontaneous chiral symmetry breaking  
in our model, changes radically the value of $\Sigma_N$, see Eq.(\ref{e:sign2}). 
We shall return to this expression in section IV.C.  
Since there are no elementary scalar 
fields in the QCD Lagrangian it is difficult to demonstrate 
how this could happen in QCD but we note that 
there are scalar {\it bound} states in QCD.\footnote{
As a simple illustration of this point one may take the example of the NJL model 
in which there are no elementary scalar mesons, but the fermion (in that 
case the constituent quark) 
$\Sigma$ term is dominated by the scalar bound state's contribution.}

\subsection{The sigma term from the scattering amplitude}

The $t$-dependent pion nucleon $\Sigma_N$ term can also be defined in terms of 
the on-mass-shell isospin symmetric amplitude as 
follows\cite{camp79,sak69,dash69}
\begin{mathletters}
\begin{eqnarray}
D^{(+)}(\nu, \nu_{B}, k_{1}^2 = m_{\pi}^2  , k_{2}^2 = m_{\pi}^2 ) 
&\equiv & 
D_{\rm PV Born}^{(+)}(\nu, \nu_{B}, k_{1}^2 = m_{\pi}^2 , k_{2}^2 = m_{\pi}^2)
+ {\Sigma_{N}(t) \over{f_{\pi}^{2}}}\; , 
\label{e:st1a} \\
D_{\rm PV Born}^{(+)}(\nu, \nu_{B}, k_{1}^2 = m_{\pi}^2 , k_{2}^2 = m_{\pi}^2)
&=&
\left({g_{\pi N}^2 \over{M}}\right) 
{\nu_{B}^{2} \over{\nu_{B}^{2} - \nu^{2}}} ~, 
\label{e:st1b} \
\end{eqnarray}
\end{mathletters}
where we have defined $D_{\rm PV Born}^{(+)}$ as given by the diagrams, 
Figs.1(a) and 1(b), using a pure pseudovector (PV) 
$\pi N$ interaction lagrangian. 
Equivalently
\begin{eqnarray}
{\tilde D}^{(+)}(\nu = 0, \nu_{B} = 0, k_{1}^2 = m_{\pi}^2 , 
k_{2}^2 = m_{\pi}^2 ) &=& {\Sigma_{N}(t = 2 m_{\pi}^2 ) \over{f_{\pi}^{2}}}
~, \
\label{e:defst}
\end{eqnarray}
which when evaluated at the unphysical Cheng-Dashen point 
gives the value of $\Sigma_N(t = 2 m_{\pi}^2)$. Here as usual 
\begin{eqnarray}
{\tilde D}^{(\pm)} &=& D^{(\pm)} - D_{\rm PV Born}^{(\pm)}
%= \left[ A^{(\pm)} + \nu B^{(\pm)}\right] - 
%\left[ A^{(\pm)} + \nu B^{(\pm)}\right]_{\rm PV Born}  
~. \
\label{e:dampl2}
\end{eqnarray}
When we compare Eq.(\ref{e:defst}) with Eq.(\ref{e:st}), we
obtain the expression 
\begin{eqnarray}
\Sigma_{N}(t) &=& \varepsilon_{3}
- M \left({m_{\pi}^2 - t - 2 \varepsilon_{2}\over{m_{\sigma}^2 - t}}\right) 
+ {\cal O}(\varepsilon^{2})~,
\nonumber \\
\Sigma_{N}(t = 2 m_{\pi}^2)
&\simeq & 
\varepsilon_{3} + 
M \left({m_{\pi}^2 + 2 \varepsilon_{2} \over{m_{\sigma}^2}}\right) 
+ {\cal O}(\varepsilon^2)~,  \
\label{e:sigma}
\end{eqnarray}
where in the last step we assume $m_\pi^2 \ll m_\sigma^2$, 
and as above we assume $m_\pi^2 \propto \varepsilon_i$, i = 1, 2, 3.
% \cite{camp79} Hamiltonian Eq.(\ref{e:chisb}), 
Eq.(\ref{e:sigma}) is in agreement with the canonical result of 
Eq.(\ref{e:sign2}) and with the original linear 
sigma model result of Campbell \cite{camp79}.

\subsection{Comparison with experiment}

The $\Sigma$ operator, Eq.(\ref{e:sigma1}), 
is often identified with the chiral
symmetry breaking ($\chi$SB) Hamiltonian itself. 
In two of the three cases in Eq.(\ref{e:dash3}),
the nucleon $\Sigma$ term is a measure of the $\chi$SB in the nucleon. 
In those cases it equals the shift of the nucleon mass $\delta M$ due 
to the $\chi$SB terms in the Hamiltonian. 
This reasoning underlies the standard interpretation
of the nucleon $\Sigma$ term as being a measure of 
the strangeness content of the nucleon \cite{kj87}. 
A large value of $\Sigma_N \simeq$ 65 MeV has often been interpreted as
a sign of a substantial $s {\bar s}$ content of the nucleon. 
We shall show that in the extended linear sigma model, Eq.(\ref{e:lag3}), 
such a large values for $\Sigma_N(t=2m_\pi^2)$ can be obtained without any strangeness 
content of the nucleon.

In the tree aproximation 
the value of the $\Sigma_N$ term in terms of the values of the 
parameters $\alpha_{i}$ of Eqs.(\ref{e:epsa}-c) is:
\begin{eqnarray}
\Sigma_N &=&
M^{0}_{N} \left(1 - \alpha_{1} - \alpha_{2}\right)
+ M \left(1 + \alpha_{2}\right)
\left({m_{\pi}^{2} \over m_{\sigma}^{2}}\right) 
+ {\cal O}(\varepsilon^2)~,
\label{e:sign3} \
\end{eqnarray}
where we use $\bar{M}^{0}_{N} \simeq 23$ MeV, $M =$ 940 MeV and 
$m_{\pi} = $ 140 MeV. 
For possible values of the $\sigma$- meson mass in the interval 
$m_{\sigma} = $ 400 - 1400 MeV\cite{u196} we have 
$M\left({m_{\pi} \over m_{\sigma}}\right)^{2}$ = 115 -  9 MeV, and hence
\begin{eqnarray}
\Sigma_N &=&
\left(1 - \alpha_{1} - \alpha_{2}\right) \times 23 {\rm MeV} +
\left(1+ \alpha_{2}\right) \times
\left(115  - 9 \right) {\rm MeV} + {\cal O}(\varepsilon^2)~.
\label{e:sign4} \
\end{eqnarray}
This range of values easily encompasses the experimentally allowed 
range of 45 - 75 MeV, for sufficiently light $m_{\sigma}$ 
and for reasonable values of $\varepsilon_i$, i = 1, 2, 3.
Note, however, that due to the large uncertainty 
in $m_{\sigma}$ this experimental value can not be used to
%determine 
effectively fix the above linear combination of the $\alpha_i$ parameters.

To compare the $\pi N$ scattering lenghts Eqs.(\ref{e:scatl+1}) 
and (\ref{e:scatl-1}) with experimental values, we discuss 
the general case $\varepsilon_i \neq 0$, $i$ = 1, 2, 3. 
\begin{eqnarray}
a_{0}^{(+)} &\simeq& 
{- g_{\pi N}^2 \over{4 \pi \left(1 + {m_{\pi} \over M}\right)}}
\left({m_{\pi} \over M}\right)\Bigg\{
\left({m_{\pi} \over2 M}\right)^{2} 
\nonumber \\
&+& g_{A}^{-2} \left[  
(1 - \alpha_{2})\left({m_{\pi}^2 \over m_{\sigma}^2 }\right)
- \alpha_{3} \left({M^{0}_{N} \over M}\right)\right] 
\Bigg\}{1 \over{m_{\pi}}}
+ {\cal O}(\varepsilon^2) ~, \
\label{e:scatl+2}
\end{eqnarray}
and 
\begin{eqnarray}
a_{0}^{(-)} &\simeq& 
{g_{\pi N}^2 \over{8 \pi \left(1 + {m_{\pi} \over M}\right)}}
\left({m_{\pi} \over M}\right)^2 
{1 \over{g_{A}^{2}}}
%\left[{1 \over{g_{A}^{2}}}+ \alpha_{3} { M^{0}_{N} \over M}
%\left(1 - g_{A}^{-2}\right) \right] 
{1 \over{m_{\pi}}} + {\cal O}(\varepsilon_3^2)~. \
\label{e:scatl-2}
\end{eqnarray}
Despite the tiny ``bare" nucleon mass $ M_N^0 \ll M$ 
the value of the isoscalar scattering length
$a_0^{(+)}$ shows significant dependence on 
both the $\alpha_{2}$ and $\alpha_3$ parameters
for values of $m_\sigma$, i.e. for $m_\sigma \leq  M_N$.
In the extended linear sigma model the theoretical value of $a_0^{(+)}$
can easily reproduce the ``old"  
experimental value $a_{0}^{(+)}|_{\rm expt.} = - 0.010(4) m_{\pi}^{-1}$,
Ref. \cite{dumb83},  and can have either sign 
with extreme values of $\alpha_{i}$ parameters. 
Recent pionic atom experiments allow for $a_{0}^{(+)}$ values
of comparable size and both signs if only
hydrogen data are taken into account\cite{leisi98}. 
The addition of the latest pionic deuteron data can flip
the sign and definitely reduces both the mean value and the 
uncertainties\cite{leisi98}. 
The new experimental value for 
$a_{0}^{(+)}|_{\rm expt.} \simeq \pm 0.0020(16) m_{\pi}^{-1}$ 
is much (almost 50 times) smaller than the ``natural" size 
obtained from the usual 
${\cal L}_{\chi SB}$ and requires 
further cancellations among these small terms. 
Thus, this latest value of $a_{0}^{(+)}|_{\rm expt.}$ 
appears to be ${\cal O}(\varepsilon^2)$.
In order that our ${\cal O}(\varepsilon^2)$ calculation 
of the $a_{0}^{(+)}$ S-wave scattering length (\ref{e:scatl+2}) 
reproduce this very small experimental value, a
very delicate cancellation between the various terms
must take place in our model that makes it very
sensitive to both 
$\alpha_2$ and $\alpha_3$ and to the value of $m_\sigma$. 
We conclude that the present tree approximation calculation is 
not sufficiently precise to 
be reliably and profitably compared with the most recent data.

To ${\cal O}(\varepsilon)$
the isospin antisymmetric scattering wave scattering length
$a_0^{(-)}$ is independent of $\alpha_i$.
%whereas to ${\cal O}(\varepsilon^2)$ it depends rather weakly 
%on the $\alpha_i$ values, ${\cal O}(\alpha^2_i)$. 
The leading order (Weinberg-Tomozawa) prediction (\ref{e:scatl-2}) 
is within one standard deviation from the (old) mean experimental 
value $a_{0}^{(-)}|_{\rm expt.} = 0.091(2) m_{\pi}^{-1}$, Ref. \cite{dumb83}, 
The new experimental value of
$a_{0}^{(-)}|_{\rm expt.} = 0.0868(14) m_{\pi}^{-1}$, Ref. \cite{leisi98},
is subject to the same caveats as for the isoscalar one described above.
%It leads, via Eq. (\ref{e:scatl-2}), to $\alpha_3 = 0.47 \pm 0.14$. 

%Another piece of information about $\alpha_3$ comes in the form 
%of the GT anomaly 
%{\bf here we are in trouble due to my wrong sign}
%\begin{eqnarray}
%g_{\pi N} f_{\pi} &=& g_{A} M (1 + \Delta_{\rm GT})
%\nonumber \\
%\Delta_{\rm GT} &=& -{\varepsilon_{3} \over{g_{A} M}} 
%= - \alpha_{3} {\bar{M}_N^0 \over{g_{A} M}} ~. \
%\label{e:deltagt}
%\end{eqnarray}
%which constrains $\alpha_{3} = - (0.97 \pm 0.49)$. 
%{\bf Please check sign again} 
%Combining this with the 
%$a_{0}^{(-)}$ value $\alpha_3 = 0.47 \pm 0.14$, we find the following region
%of overlap $0.5 \leq \alpha_3 \leq 0.61$. 
% and then from the nucleon 
%$\Sigma$ term Eq. (\ref{e:sign4}) follows $\alpha_{2} \simeq 0$.
%modification of the linear sigma model leads to a value of $a_{0}^{(+)}$
%in better agreement with experiment.

%The nucleon $\Sigma$ term  determines a linear 
%combination of the two parameters $\alpha_{2,3}$, Eq. (\ref{e:sign4}). With
%$m_{\sigma}$= 600 MeV, one finally finds $\alpha_1 = 0.45 \pm 0.23$ and
%$0 \leq \alpha_3 \leq 0.18$. 
%{\bf These numbers not checked}. 
%These values ought to be taken {\it cum grano salis} because of all the 
%{\it caveats} stated earlier in the text. 

\section{Relationship to chiral perturbation theory}

A 
%Dyson-Foldy-Weinberg 
``chiral rotation'' defined in
the limit $m_{\sigma} \to \infty$ by
\begin{mathletters}
\begin{eqnarray}
N &=& \sqrt{\cal R} 
\left( 1 + i \gamma_{5} \bbox{\tau} \cdot \bbox{\xi}\right)\psi \\
\bbox{\pi} &=& {\cal R} \bbox{\phi} \\
\sigma &=& f_{\pi} {\cal R} \left(1 - \bbox{\xi}^{2} \right) \\
{\cal R} &=& 
\left[1 + \left({\bbox{\phi} \over{2 f_{\pi}}}\right)^{2} \right]^{-1} 
= \left[1 + \bbox{\xi}^{2} \right]^{-1}~,
\label{e:rot} \
\end{eqnarray}
\end{mathletters}
leads, by way of  standard arguments \cite{wein67,walecka95} from 
the linear sigma model 
Lagrangian without the extra derivative interaction terms in Eq.(\ref{e:lag3}), 
to the nonlinear one 
\begin{eqnarray}
{\cal L} - {\cal L}_{\chi SB} &=& 
\bar{N}[{\rm i}{\partial{\mkern-10mu}{/}} - M ]N +
{1 \over 2} 
%{\cal R} \left[
{\cal R}^2 \left(\partial_{\mu} \bbox{\phi}\right)^{2} 
%- m_{\pi}^{2} \bbox{\phi}^{2} \right] 
\nonumber \\
&+& 
{\cal R} \left({1 \over{2 f_{\pi}}}\right)
\left(\bar{N}\gamma_{\mu} \gamma_{5} \bbox{\tau}N 
\right) \cdot \partial^{\mu} \bbox{\phi} 
\nonumber \\
&-& {\cal R}
%\left({g_{\pi NN} \over{2 g_{A} M}}\right)
\left({1 \over{2 f_{\pi}}}\right)^{2} 
\left(\bar{N}\gamma_{\mu} \bbox{\tau}N \right) \cdot
\left(\bbox{\phi} \times \partial^{\mu} \bbox{\phi}\right) ~.
\label{e:lag2} \
\end{eqnarray}
The above form of the nonlinear Lagrangian (\ref{e:lag2}) 
differs from Weinberg's\cite{wein67} by the absence of an 
{\it ad hoc} factor $g_A$ in front of the ``pseudovector'' coupling term.
The source of this difference, as emphasized by Weinberg himself, 
was the need to have both the empirical $g_A$ factor in 
the axial current and the correct two-pion-nucleon contact interaction.
We shall now show that the extended linear sigma model, Eq.(\ref{e:lag3}), 
leads to Weinberg's nonlinear 
sigma model Lagrangian, i.e., that the extra 
terms in Eq.(\ref{e:lag3}) promoted by Bjorken-Nauenberg and by Lee provide
precisely the difference prescribed {\it ad hoc} by Weinberg. 
The extra terms in Eq.(\ref{e:lag3}) can be written in terms of the 
currents ${\bf V}_{\mu}$, ${\bf v}_{\mu}$ and
${\bf A}_{\mu}$, ${\bf a}_{\mu}$:
% of Eq.(\ref{e:lag2}):
\begin{eqnarray}
{\cal L}_{\rm bn} &=&
\left({g_{A} - 1 \over{f_{\pi}^{2}}}\right) \left[
\left(\bar{\psi}\gamma_{\mu} {\bbox{\tau}\over 2}\psi \right) \cdot
\left(\bbox{\pi} \times \partial^{\mu} \bbox{\pi}\right)
+ \left(\bar{\psi}\gamma_{\mu} \gamma_{5} {\bbox{\tau}\over 2}\psi 
\right) \cdot
\left(\sigma \partial^{\mu} 
\bbox{\pi} - \bbox{\pi} \partial^{\mu} \sigma \right) \right]
\nonumber \\
&=& 
\left({g_{A} - 1 \over{f_{\pi}^{2}}}\right) \left[
{\bf V}_{\mu} \cdot {\bf v}^{\mu} + {\bf A}_{\mu} \cdot {\bf a}^{\mu}\right] ~.
\label{e:bn} \
\end{eqnarray}
%The nucleon and mesonic parts of 
%The Noether  current  of Eq.(\ref{e:lag2}) is the sum of 
The ${\bf V}_{\mu}^{a}$ and ${\bf v}_{\mu}^{a}$ are
\begin{eqnarray}
{\bf V}_{\mu}^{a} &=&
%\left(\bar{\psi}\gamma_{\mu} {\bbox{\tau}\over 2}\psi \right)^{a} = 
{\cal R} 
\Bigg\{ \left(1 - \bbox{\xi}^2 \right)
\left(\bar{N}\gamma_{\mu} {\bbox{\tau} \over 2}N \right)^{a} 
\nonumber \\
&-& 
\bar{N} \gamma_{\mu} \gamma_{5} 
\left(\bbox{\tau} \times \bbox{\xi}\right)^{a}N + 
\bbox{\xi}^{a} \bar{N} \gamma_{\mu} 
\left(\bbox{\tau} \cdot \bbox{\xi} \right) N  \Bigg\} 
\label{e:vec1} \\
{\bf v}_{\mu}^{a} &=&
%{\left(\bbox{\pi} \times \partial_{\mu} \bbox{\pi}\right)^{a} =
{\cal R}^2 
\left(\bbox{\phi} \times \partial_{\mu} \bbox{\phi}\right)^{a} ~.
\label{e:vec2} \
\end{eqnarray}
Similarly, 
%the two parts of the Eq.(\ref{e:lag2}) axial-vector Noether current are:
%{\bf compare these to Eq.(\ref{e:axi4})}
\begin{eqnarray}
{\bf A}_{\mu}^{a} &=&  
%\left(\bar{\psi}\gamma_{\mu}\gamma_{5} {\bbox{\tau}\over 2}\psi \right)^{a}= 
{\cal R} 
%\left(1 - \left({\bbox{\pi} \over 2 f_{\pi}}\right)^{2} \right)
\Bigg\{ \left(1 - \bbox{\xi}^2 \right)
\left(\bar{N}\gamma_{\mu}\gamma_{5} {\bbox{\tau}\over 2}N \right)^{a} 
\nonumber \\
&-& 
%\left({1 \over{f_{\pi}}}\right)   
\bar{N} \gamma_{\mu}  
\left({\bbox{\tau} \times \bbox{\xi}}\right)^{a}N 
+ \bbox{\xi}^{a} \bar{N} \gamma_{\mu} \gamma_{5} 
\left(\bbox{\tau} \cdot \bbox{\xi} \right) N  \Bigg\} 
\label{e:ax1} \\
{\bf a}_{\mu}^{a} &=& 
{\cal R}^2 f_{\pi} \left[
\partial^{\mu}\bbox{\phi}^{a} \left(1 - \bbox{\xi}^2 \right)
+  2 \bbox{\xi}^{a} 
\left(\bbox{\xi} \cdot \partial^{\mu}\bbox{\phi}\right) \right] ~.
\label{e:ax2} \
\end{eqnarray}
Inserting these into Eq. (\ref{e:bn}) we find
\begin{eqnarray}
{\cal L}_{\rm bn} &=& \left(g_{A} - 1 \right) {\cal R} 
\left(\bar{N}\gamma_{\mu} \gamma_{5} \bbox{\tau}N 
\right) \cdot \partial^{\mu} \bbox{\xi} ,
\label{e:bn1} \
\end{eqnarray}
which when combined with Eq.(\ref{e:lag2}) leads to Weinberg's 
nonlinear Lagrangian with $g_A \ne 1$. 
One can now write the resulting nonlinear 
Lagrangian in the notation of chiral perturbation theory and thus convince 
oneself that this is equivalent to the lowest order Lagrangian of Gasser,
Sainio and \v Svarc [GS\v S] \cite{gss88}. 
Conversely, one should be able to convert finite-chiral-order terms in 
the GS\v S nonlinear chiral Lagrangian into extended linear ones.
%The extended linear sigma model
%now has derivative-coupled terms that allow chiral (power) counting to be 
%performed just as in the nonlinear chiral Lagrangians \cite{wein79}.
This is more than an academic point, for it makes it clear that the choice 
between the linear and nonlinear realizations is a matter of convenience.
Quite often it is more expedient to work in the representation wherein
one has the $\sigma$, or $s$ fields from the beginning, rather than building 
it up from the pions. 
Moreover, the linear lagrangian is always a polynomial in the meson fields,
rather than a fractional, or even (transcedental) exponential  function of 
$\pi$ as is the case in the nonlinear realization. 

\section{Summary and Conclusions}

In this work we have shown that the extension of the linear $\sigma$ 
model allows $g_{A} \neq 1$ in the axial current in the linear 
realization of chiral symmetry.
The chiral charge algebra holds in the 
extended linear sigma model 
despite the fact that the \underline{spatial} part of the nucleon axial 
current is 
renormalized by $g_A$, because the nucleon axial charge is not renormalized. 

We evaluated the elastic $\pi$N scattering amplitude in the tree 
approximation with three kinds of $\chi$SB terms similar to Ref.\cite{camp79}. 
The $a_{0}^{(-)}$ scattering length is now in agreement with the
Weinberg-Tomozawa 
result, and we can obtain a small  $a_{0}^{(+)}$ scattering length value 
in contrast to  
the original linear sigma model.

The $\Sigma_N$ term 
with three different $\chi$SB terms was also evaluated.
In the tree aproximation 
the $\Sigma_N$ term from the canonical operator definition
using the equations of motion  
coincides with the result derived from the $\pi$N scattering
amplitude. The vacuum
matrix element of the $\Sigma$ operator puts one constraint on 
a linear combination of the three different $\chi$SB parameters 
$\varepsilon_{i}$, $i$ = 1, 2, 3. 
%{\bf We find that 
%the $\Sigma$ term matrix elements differ from one approximation 
%to another within the same model. WHAT DO YOU MEAN???}
It is noteworthy that in  
our extended linear sigma model a large value for $\Sigma_N$ can easily be
obtained without any $s \bar{s}$ components in the nucleon. The reason for 
this 
is that the scalar $\sigma$ meson can make a large contribution 
to $\Sigma_N$ depending on the value of the mass $m_{\sigma}$.

Finally we showed that a chiral rotation of the extended linear sigma model 
Lagrangian leads to the lowest-order $\pi$N $\chi$PT Lagrangian in
the limit $m_{\sigma} \to \infty$. 

We close with several suggestions for future research:
(i) Derive $\varepsilon_i$, for $i$ = 1, 2, 3 from quark models or QCD [for a
sketch of such a derivation in the NJL model, see Appendix \ref{app2}].
(ii) Apply the extended sigma model to a re-evaluation of a 
possible pion condensation in nuclear matter, where $g_{A} \neq 1$ 
is very important, but has not been consistently implemented to date.
(iii) 
%Use the extended sigma model to investigate possible (model dependent)
%values of the low energy constants in the $\chi$PT lagrangian. 
Establish a relation between the free parameters of the extended 
linear sigma model and 
%to investigate possible (model dependent) values of the 
the low-energy constants in the $\chi$PT lagrangian 

\acknowledgments
This work was supported in parts by an NSF grant no. PHY-9602000.

\appendix
\section{Dashen's relation at the one-loop level}
\label{app}

Here we follow Section V of Ref.\cite{ax96} 
and show that at the one-nucleon-loop level we can derive Eq.(\ref{e:dash4}). 
[Analogous calculations at the one-meson-loop self-consistent approximation 
level can be performed along the lines of Ref.\cite{dms96}.] 

The Hartree + RPA approximation
can be defined by three Schwinger-Dyson integral equations:
(i) the zero-body or vacuum equation, 
(ii) the one-body or the fermion mass gap equation, 
and (iii) the two-body or one-meson Bethe-Salpeter equation, shown
in Figs.~5.a, b, and c of Ref. \cite{ax96}, respectively. 
The Bethe--Salpeter equation for 
the $N\bar{N}$ pseudoscalar scattering amplitudes 
is separable and has as an exact 
solution in  Hartree+RPA the following expression:
\begin{eqnarray}
 D_{\pi}(k) &=& {1 \over  k^2 - 
\Sigma_{\pi}^{(\rm RPA)}(k)} \ ,
\label{e:prop}
\end{eqnarray}
where $\Sigma^{(\rm RPA)}(k)$ is a sum of a single one-nucleon-loop 
polarization diagram plus one ``tree" diagram.
%, see Fig.~5.c. 
%Note that the above solution is just a geometric series in 
%$\Sigma^{(\rm RPA)}(k)$.
%The Hartree nucleon self-energy $\Sigma^{(\rm H)}$ is momentum independent,
%but the RPA 
%scalar $\Pi_{\sigma}^{(\rm RPA)}(k)$ and 
%pseudoscalar $\Pi_{\pi}^{(\rm RPA)}(k)$ polarization function depends 
%on the four-momentum squared $k^2$.
%, or one loop approximation, satisfy the following 
The Schwinger-Dyson equations now read ($v$ = $f_\pi$)\cite{ax96}
%\mediumtext
\begin{mathletters}
\begin{eqnarray}
v &=& - {\varepsilon_{1} \over \mu_{0}^2} 
+ \lambda_{0} {v^{3} \over \mu_{0}^2} + {i \over \mu_{0}^2}
g_{0} N_f \intop\limits\hskip -1pt{d^4 p \over (2\pi)^4}{4 M \over
p^2 - M^2}
\label{e:vac}\\
%\Sigma^{(\rm H)} &=&
M &=& M_{0} + g_{0} v = \varepsilon_{3} + g_{0} v
\label{e:gap} \\
\Sigma_{\pi}^{(\rm RPA)}(k) &=& 
2 \varepsilon_{2}
- \mu_{0}^2 + \lambda_{0} v^{2}
 + g_{0}^{2} \Pi_\pi^{(\rm RPA)}(k)
\label{e:rpap} \
\end{eqnarray}
\end{mathletters}
%\narrowtext
where Eq.(\ref{e:gap}) is the same as Eq.(\ref{e:na}). 
%nothing but the (approximate)
%non-chiral Goldberger-Treiman (GT) relation, which becomes exact in the chiral 
%limit (here that means vanishing bare nucleon mass 
%$\varepsilon_{3} = M_0 = 0$).
%Several comments are in order here. 
The pion polarization function $\Pi_\pi^{(\rm RPA)}(k)$ can be written as
\mediumtext
%\begin{mathletters}
\begin{eqnarray}
%{-{\rm i}}
\Pi_\pi^{(\rm RPA)}(k) &=&
4 i N_f \intop\limits\hskip -1pt{d^4 p \over (2\pi)^4} {1\over p^2- M^2}
- 2 i N_f k^2 I(k)
%\label{e:pipi} 
\nonumber 
\\
&=&
{1 \over M} \langle {\bar \psi}\psi \rangle_{0} - 2 i N_f k^2 I(k)~,
\label{e:pisig} \
\end{eqnarray}
%\end{mathletters}
\narrowtext
where we introduced the logarithmically divergent integral
\begin{eqnarray}
I(k) &=&
\intop\limits\hskip -1pt{d^4 p \over (2\pi)^4}
{1\over [p^2 - M^2][(p+k)^2 - M^2]} \ .
\label{e:i}
\end{eqnarray}
In order to prove the Dashen relation (\ref{e:dash4})
% away from the chiral limit ($M_0 \neq 0$), 
we rewrite Eq. (\ref{e:vac}) using Eq.(\ref{e:pisig}) as follows
\mediumtext
\begin{eqnarray}
%1 &=& - {\varepsilon_{1} \over{v \mu_{0}^2}} + 
%\lambda_{0} {v^{2} \over \mu_{0}^2} + {i \over \mu_{0}^2}
%4 g_{0} {M \over v} N_f 
%\intop\limits\hskip -1pt{d^4 p \over (2\pi)^4}{1\over p^2 - M^2}
%\nonumber \\
- \mu_{0}^2 + \lambda_{0} v^{2} &=& 
{\varepsilon_{1} \over{v}}
- {g_{0} \over v}  \langle {\bar \psi}\psi \rangle_{0} ~. \
\end{eqnarray}
\narrowtext 
When we compare this equation with the tree approximation results, Eq.(\ref{e:nc}), 
we see that the last term above is beyond the tree-level.  
Insert this into  Eq.(\ref{e:rpap}) to find to lowest order in $\varepsilon_i$ 
(as $k \rightarrow 0$): 
%\begin{mathletters}
\begin{eqnarray}
m_\pi^{2} =
\Sigma_{\pi}^{(\rm RPA)}(0) &=&  
2 \varepsilon_{2} - \mu_{0}^2 + \lambda_{0} v^{2}
+ {g_{0}^{2} \over M} \langle {\bar \psi}\psi \rangle_{0}
\nonumber \\ 
&=& 
{\varepsilon_{1} \over{v}}
- {g_{0} \over v}  \langle {\bar \psi}\psi \rangle_{0}
+ 2 \varepsilon_{2} 
%- \mu_{0}^2 + \lambda_{0} v^{2}
+ {g_{0}^{2} \over M} \langle {\bar \psi}\psi \rangle_{0}
\nonumber \\ 
&\simeq& 
{\varepsilon_{1} \over{v}} + 2 \varepsilon_{2} 
- {\varepsilon_{3} \over v^{2}} \langle {\bar \psi}\psi \rangle_{0}
\label{e:pmass}~, \
\end{eqnarray}
%\end{mathletters}
%{\bf You assumed lim $k^2 I(k) \to 0$ as $k \to 0$.}
where we used the GT relation (\ref{e:gap}).
Equation (\ref{e:pmass}) is equivalent to Eq.(\ref{e:dash4}) to leading 
order in $\chi$SB parameters.
Thus we have demonstrated the necessity of a self-consistent gap equation for
the validity of Dashen's formula when $\chi$SB is determined by Eq.(\ref{e:chisb}).

\section{Sketch of a derivation of $\varepsilon_{1}$ and $\varepsilon_{3}$}
\label{app2}

We shall use the bosonization technique in a simple chiral quark (NJL) model to 
show that $\varepsilon_{1}$ is related to $\varepsilon_{3}$ 
at the quark level. This is just a sketch 
meant to illustrate an approach to the more challenging case of nucleons.

The NJL model Lagrangian density is 
\begin{eqnarray}
{\cal L}_{\rm NJL} &=& 
\bar{\psi}\left[i {\partial{\mkern-10mu}{/}} - m^{0} \right]\psi 
%- g_{0}\bar{\psi}\left[\sigma + 
%i \gamma_{5} \bbox{\pi} \cdot \bbox{\tau} \right]\psi  
\nonumber \\
&+& 
G \left[\left(\bar{\psi}\psi \right)^{2}
+ \left(\bar{\psi} i \gamma_{5} \bbox{\tau} \psi\right)^{2} \right] 
\label{e:njl} \
%+ {\cal L}_{\chi SB} \
\end{eqnarray}
The following substitution
\begin{mathletters}
\begin{eqnarray}
- g_{0} \sigma 
&=& 
G \left(\bar{\psi}\psi \right) 
\label{e:ida} \\ 
- g_{0} \bbox{\pi}
&=& 
G \left(\bar{\psi} i \gamma_{5} \bbox{\tau}\psi \right) 
\label{e:idb}~, \
\end{eqnarray}
\end{mathletters}
for one of the two quark bilinears
leads to the (semi-bosonized) linear $\sigma$ model interaction Lagrangian
\begin{eqnarray}
{\cal L}_{\rm int} &=& 
%\bar{\psi}\left[i {\partial{\mkern-10mu}{/}} - m^{0} \right]\psi 
- g_{0} \bar{\psi}\left[\sigma + 
i \gamma_{5} \bbox{\pi} \cdot \bbox{\tau} \right]\psi  
\label{e:lsm}\
\end{eqnarray}
The chiral symmetry breaking current quark mass term is 
\begin{eqnarray}
{\cal L}_{\chi SB} &=& 
- m^{0} \bar{\psi}  \psi 
= - \varepsilon_{3} \bar{\psi} \psi
\nonumber \\ 
&=& 
m^{0} {g_{0} \over G} \sigma = \varepsilon_{1} \sigma . 
\label{e:csb} \
\end{eqnarray}
Note that Eq. (\ref{e:ida}) implies 
(using the linear sigma model relations)
\begin{eqnarray}
- g_{0} \langle \sigma \rangle_{0} 
&=& 
G \langle {\bar \psi}\psi \rangle_{0} = - g_{0} f_{\pi} = - m
\label{e:vev}~. \
\end{eqnarray}
This in turn leads to 
%{\bf check signs}
\begin{eqnarray}
\varepsilon_{1}  
&=& 
\varepsilon_{3} {g_{0} \over G} = - {m^{0} \over f_{\pi}} 
\langle {\bar \psi}\psi \rangle_{0} = m_{\pi}^{2} f_{\pi}
\label{e:rel}~, \
\end{eqnarray}
where the last step follows from the GMOR relation, which can be explicitly 
demonstrated in the NJL model at the quark level. 

Chiral symmetry breaking coefficients have been calculated at the mesonic level
in a more sophisticated chiral quark (``global color'') model in Ref. 
\cite{meis98}. 
%{\bf what about ENJL calculations of J.H. Bijnens et al.?}
The challenge is to extend this analysis to the nucleon case. 
This can presumably be done by solving the three-quark Faddeev-Bethe-Salpeter
equation, see Ref.\cite{alkof98}, and  
calculate $\varepsilon_{3}$ at the nucleon level.

\widetext
%\begin{figure}
\input{axodraw.sty}
%
% ***********************************************************
%\begin{figure}
\begin{figure}
\caption{ 
      The elastic $\pi$N scattering amplitude: (a) the direct-, and 
%kernel (relativistic potential) of the BSe
%in the one-boson-exchange (OBE) approximation 
(b) crossed nucleon-pole diagrams; (c) the contact, or sea-gull diagram, 
and (d) the $\sigma$-meson-pole diagram.
      The dashed line denotes a pion; the zig-zag line
      denotes a $\sigma$ meson, the solid line denotes a nucleon.
% and the double solid line in (a) represents a deuteron.
% The hatched ``blob" in (a) represents the deuteron wave function, and
% the box represents the relativistic potential.
      \label{fig1}}
\end{figure}

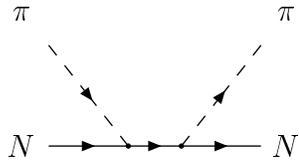
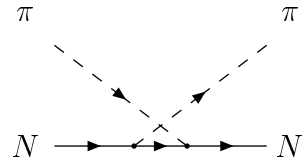
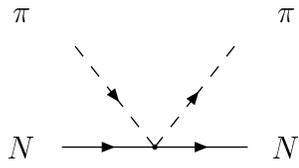
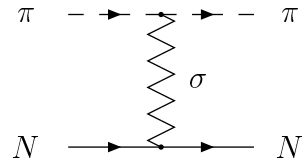
\end{document}